\RequirePackage{fix-cm}
\documentclass[natbib,smallextended]{svjour3}       
\smartqed  
\usepackage{graphicx}
\usepackage{color}
\usepackage{ulem}
 \journalname{my journal}
\newcommand{\ikoma}{\textcolor{black}}

\newcommand{\sakujo}[1]{\if0{#1}\fi}

\begin{document}

\title{Water Partitioning in Planetary Embryos and Protoplanets with Magma Oceans 
}


\author{M. Ikoma \and L. Elkins-Tanton \and K. Hamano \and J. Suckale}


\institute{
	M. Ikoma \at
	\textit{Department of Earth and Planetary Science, The University of Tokyo, Tokyo, Japan; also, Research Center for the Early Universe (RESCEU), The University of Tokyo, Tokyo, Japan}
	\and
	L. Elkins-Tanton \at 
	\textit{School of Earth and Space Exploration, Arizona State University, Arizona, USA}
	\and
	K. Hamano \at
	\textit{Earth-Life Science Institute, Tokyo Institute of Technology, Tokyo, Japan}
	\and
	J. Suckale \at 
	\textit{Department of Geophysics, Stanford University, California, USA}
	}

\date{Received: date / Accepted: date}

\maketitle

\begin{abstract}
The water content of magma oceans is widely accepted as a key factor that determines whether a terrestrial planet is habitable. 
Water ocean mass is determined as a result not only of water delivery and loss, but also of water partitioning among several reservoirs. 
Here we review our current understanding of water partitioning among the atmosphere, magma ocean, and solid mantle of accreting planetary embryos and 
protoplanets just after giant collisions. 
Magma oceans are readily formed in planetary embryos and protoplanets in their accretion phase. 
Significant amounts of water are partitioned into magma oceans, provided the planetary building blocks are water-rich enough. 
Particularly important but still quite uncertain issues are how much water the planetary building blocks contain initially and how water goes out of the solidifying mantle and is finally degassed to the atmosphere. 
Constraints from both solar-system explorations and exoplanet observations and also from laboratory experiments are needed to resolve these issues.
\end{abstract}

\section{Introduction}
\label{intro}
Water plays crucial roles in the climate, plate tectonics, mantle dynamics, and thermal evolution of terrestrial planets. 
This is partly because water vapor is a strong greenhouse gas. 
Also, water makes a great contribution to control the CO$_2$ partial pressure and thus the surface temperature of the Earth through continental weathering, which occurs by reactions between water and silicate/carbonate \citep{Walker+81}. 
Furthermore, on the Earth, water is known to \sakujo{have} influence \sakujo{on} the lithospheric density and viscosity, which likely leads to plate tectonic \citep[e.g.,][]{Korenaga13}.
Likewise, water affects the rheology of mantle materials and thus the thermal evolution of terrestrial planets.

The amount of water in the hydrosphere is thought to be a key factor that determines the surface environment (\ikoma{and} habitability) of terrestrial planets. 
The total mass of the Earth's oceans accounts for only 0.023~wt\% of the Earth's total mass. 
Such a small amount of water allows continents to be present on the Earth. 
If seawater were about three times more abundant than at present \citep[e.g.,][]{Maruyama+13}, there would be no continents. 
If so, continental weathering would hardly work and, consequently, the Earth's climate would be much different from the present \citep{Abbot+12}. 
Also, the evolution and lifetime of habitable planets is known to differ greatly, depending on the amount of seawater \citep{Abe+11,Kodama+15}.

Of great importance for understanding the formation, evolution, and habitability of terrestrial planets are, hence, not only water delivery and loss, but also water partitioning among the reservoirs such as the atmosphere/oceans, crust, mantle, and core. 
This is the main focus of this chapter. 
\ikoma{Water budget of the Earth is also discussed in Chapter~4.} 
Delivery and loss processes of water are mainly discussed in Chapter~\ikoma{10} and in Chapters~\ikoma{13} and \ikoma{14}, respectively. 
However, one can never consider water partitioning separately from the delivery (or accretion) and loss processes, because all those processes sometimes occur concurrently. 
Water moves into and from molten portions of a mantle, namely, magma oceans. 
Magma oceans are formed because of the large amount of energy brought by accretion of solids. 
Also, magma oceans are maintained by the strong blanketing effect of atmospheres, which accrete simultaneously with the growth of planetary embryos and are lost by intense stellar irradiation. 

The key questions that we address in this chapter are, thus,  
\begin{enumerate}
	\item[(1)] how magma oceans form and solidify and how deep are they?
	\item[(2)] how much water is dissolved from atmospheres into magma oceans?
	\item[(3)] how \ikoma{is} water transported in the interior of solidifying magma oceans and how the transported water goes out to atmospheres?
	\item[(4)] how much water is finally retained in the hydrosphere and interior of planetary embryos and planets?
\end{enumerate}
Note that partitioning of hydrogen and oxygen into the core is potentially of considerable importance \citep[e.g.,][]{Rubie+04,Rubie+15}. However, while hydrogen is inferred to be contained in the core \citep[e.g.,][]{Nomura+14}, it still remains quite uncertain, because partitioning of hydrogen into liquid Fe is extremely difficult to study experimentally \citep[e.g.,][]{Okuchi97}. Detailed discussion about water partitioning into the core is beyond the scope of this chapter.

The rest of this chapter is organized as follows. 
We first outline theories regarding the growth of planetary embryos and the formation and depth of magma oceans of different types in Sect.~\ref{sec:12-2}. 
In Sect.~\ref{sec:12-3}, we discuss the accumulation of vapor atmospheres and water partitioning between the atmosphere and magma ocean in accreting planetary embryos. 
In Sect.~\ref{sec:12-4}, we discuss the degassing of water in the solidifying mantles and the secondary formation of atmospheres on protoplanets after giant collisions. 
Finally we present conclusions of this chapter in Sect.~\ref{sec:12-6}.

\section{Magma Oceans on Growing Planetary Embryos}
\label{sec:12-2}


\subsection{Growth of planetary embryos}
\label{sec:12-2-1}

Planetary embryos are thought to experience a great number of collisions with objects of different sizes, ranging from 1~cm to 1000~km, in the accretion stages.
First there would be no doubt that planet formation occurs in a disk-shaped gas that surrounds the central proto-star, which is in general called a protoplanetary disk or the solar nebula (or sometimes the proto-solar disk) in the case of the solar system. 
Observation of young stellar clusters indicates that protoplanetary disks contain dust particles of $\mu$m size. 
Such small particles are dynamically coupled with the ambient disk gas and, thus, are unable to accrete directly to form planets. 
Because of their Brownian motion, dust particles meet each other and stick together to grow to a certain size. They then start accreting by mutual collisions after being slightly decoupled from the ambient gas.
The size of the initial building blocks for solid planets, however, remains a matter of debate.

Classical theories for the formation of the solar system assume that the initial building blocks, 
which are called \textit{planetesimals}, are about 10~km in size (corresponding to about 10$^{15}$~kg) around the Earth's heliocentric distance. 
This is based on the gravitational instability model for planetesimal formation \citep[][]{Safronov72,Hayashi72,Goldreich+Ward73}, 
where vertical settling of dust particles forms a thin layer with high solid concentration, followed by gravitational fragmentation of the layer.  
The thickness of the layer determines the typical planetesimal size. 
Several mechanisms are, however, known to hinder the formation of such thin dust layers in protoplanetary disks. 
Instead of vertical settling, radial motion of dust particles may result in areas with a high dust concentration through turbulent concentration and/or streaming instability \citep[see][for a recent review]{Johansen+14}. 

Current understanding of planetary accretion from planetesimals is as follows. 
First, planetesimals grow in a relatively orderly fashion through mutual collisions. 
When planetesimal mass is roughly 10$^{20}$~kg, which is similar to the mass of Vesta, 
gravitational focusing becomes so effective that the largest planetesimal starts to grow much faster than the other smaller ones in a runaway fashion \citep{Wetherill+Stewart89, Kokubo+Ida96}. 
In this runaway growth phase, a bimodal size distribution of solids develops, consisting of a small number of large bodies, which are called planetary embryos, and a large number of small planetesimals. 
In this phase, planetary embryos experience successive collisions with small planetesimals. 
The runaway growth is, however, known to end relatively soon, 
because planetary embryos stir the surrounding planetesimals by gravitational scattering and enhance their random velocities, thus lowering the gravitational focusing effect \citep{Ida+Makino93}. 
Subsequently, several large embryos grow in an orderly fashion. 
This is called the oligarchic growth stage \citep{Lissauer87,Kokubo+Ida98}.
Those embryos finish their growth after capturing almost all the planetesimals in their own feeding zones with a width of about 10 Hill radii. 
The final mass of those embryos, which is called the isolation mass, depends on the initial solid surface density. 
In the minimum mass solar nebula \citep[][MMSN, hereafter]{Hayashi81}, the isolation mass is almost equal to the Mars mass. 
Further growth requires collisions between those isolated embryos, which are often called giant collisions. 
This means that the Earth and Venus likely experienced several giant collisions. 
In the case of the Earth, the final collision is believed to be the moon-forming one \citep[e.g.,][]{Hartmann+Davis75, Benz+86, Ida+97}.  

The timing of disk gas dispersal has a great influence on the volatile contents, magma ocean formation, and redox state of planetary embryos and planets (see Sect.~\ref{sec:12-3}). 
Also, it affects the dynamics and migration of planetary embryos. 
Unfortunately, we have no direct evidence for the timing of the solar nebula dispersal. 
Instead, astronomical observations of young stellar clusters indicate that protoplanetary disks dissipate in $\sim$5--10 million years \citep[e.g.][]{Haisch+01}. 
On the other hand, according to recent Hf-W chronology, 
the formation ages of Mars and the Earth are about a few million \citep{Dauphas+11} and $>$ 30 million years \citep{Halliday+14}, respectively. 
Thus, Earth formation is likely to have been completed well after solar nebula dispersal, 
whereas Mars may have formed while the solar nebula still existed. 
Note that some isotopic ratios, such as D/H and $^{18}$Ne/$^{20}$Ne, of materials from the lower mantle of the Earth suggest that the Earth gained the nebular components at some point during its formation \citep[e.g.,][]{Hallis+15}. 
Thus, although not conclusive, a possible scenario in the framework of the above classical formation model is that the nebular gas existed until Mars-size protoplanets were formed and its dissipation took place during the giant collision stage.

Recently, there has been a movement to make a drastic re-examination of planet formation theories based on planetesimal accretion. 
This is motivated by the well-known theoretical difficulty in forming gas giants before disk gas dispersal. 
Within the classical framework based on planetesimal accretion, 
in a protoplanetary disk with a normal surface density like MMSN, 
the in-situ core growth proceeds so slowly at $>$ 5 AU that runaway gas accretion never occurs before disk gas dispersal \citep[e.g.,][]{Pollack+96,Ikoma+00}. 
A promising way to hasten the core growth is to consider much smaller objects of $\sim$ 1-10~cm in size, which are often termed pebbles. 
This is because such objects are subject to gas drag, which keeps their velocity dispersion low and enlarges their cross section for collision with proto-cores surrounded by envelopes  \citep{Inaba+03a, Inaba+03b, Ormel+10}. 
Indeed, recent planet formation theories in which pebbles are assumed to accrete to earlier-formed planetary embryos have demonstrated successfully that cores of critical mass are formed well before disk dispersal \citep{Lambrechts+12, Lambrechts+14a}. 
In such theories, radial drift of pebbles plays an important role in supplying materials. 
This could have a great influence also on terrestrial planet formation and water delivery \citep{Morbidelli+15}. 
The pebble accretion paradigm is, however, still developing. 
There is no detailed quantitative argument regarding incorporation and partitioning of water in planetary embryos and planets in the framework of the pebble accretion paradigm.  
Thus, our discussion below is based only on planetesimal accretion, unless otherwise stated.

\subsection{Origins of magma oceans on planetesimals and planets}
\label{sec:12-2-2}
Three heat sources may dominate the period of planetary accretion and solidification \citep[e.g.,][]{Elkins-Tanton12,Rubie+15}:
\begin{enumerate}
\item[(1)] \textit{Impact}---The heat from accretionary impacts primarily affects the near-surface and surface of the planet. The energy involved in impacts between planetesimals is likely insufficient to significantly melt those small bodies; impact-induced melting becomes more important in planetary embryos, and dominant in planets \citep[e.g.,][]{Davison+10}.
\item[(2)] \textit{Differentiation}---Metal-silicate differentiation may occur at mid-mantle depths in a magma ocean due to local differentiation between heavy metal and light silicates \citep{Sasaki+Nakazawa86,Tonks+Melosh92} or closer to the core in larger impacts or as the metal phase gradually sinks towards the center of the impacted protoplanet \citep{Monteux+09}.
In all the cases the potential energy conversion to heat likely primarily affects the deep mantle and core. Again this heat source is likely important for melting only in planetary embryos and planets. 
\item[(3)] \textit{Radiogenic decay}---Heat from radiogenic elements may be evenly dispersed unless fractional solidification of silicate liquids concentrates them. 
Following fractional solidification of a silicate mantle on a planet of Mars-sized or smaller, the incompatible elements, including almost all the radiogenic elements, will be concentrated in the last dregs of liquid near the surface of the planet. This material is compositionally dense and likely to sink to the bottom of the mantle, possibly leaving small pockets behind near the surface of the planet. On a larger planet, both a basal and an upper magma ocean may fractionate separately, though with the same outcome that the incompatible elements will be concentrated at the bottom in the end \citep{Labrosse+07,Brown+14}. On large planets, therefore, heating may occur simultaneously from above and below. In planetesimals, melting is almost certainly the near-exclusive domain of the short-lived radioisotope $^{26}$Al, since the small mass of these bodies precludes significant melting from core formation or impacts.
\end{enumerate}

Different heat sources may translate into different types of magma oceans, 
from (1) magma ponding at the surface of a largely solid planet, 
to (2) a basal magma ocean overlain by solids, 
to (3) a whole-mantle magma ocean. 
Magma oceans on planetary surfaces are mostly created by impactors. 
The shock pressure is nearly uniform in a spherical region around the impact, the isobaric core, and decays rapidly with distance \citep{Croft82, Pierazzo+97}. 
In the isobaric core, the kinetic energy of the impact increases the temperature by several hundred degrees depending mostly on the properties of the impactor \citep{Monteux+07}. 
Surface magma oceans of different sizes are likely to have existed on the early Earth at different stages of its evolution due to the varying energies of accretionary collisions \citep{Chambers13} and the late heavy bombardment \citep{Chou78}.

A basal magma ocean could form in one of two ways: First, it could have resulted from a density inversion. 
The compositionally dense material remaining at the end of fractional solidification could either sink as a liquid or as a solid that melts through the heat of long-lived radiogenic isotopes \citep{Brown+14}. 
This possibility was explored in detail for the early Earth by \citet{Labrosse+07}, who concluded that fractional solidification was important for producing deep, dense material that still might exist today. 
Second, a basal magma ocean may be created by the potential energy of core formation. 
For the early Earth, the release of gravitational energy during core formation may induce melting of the whole planet \citep{Ricard+09}, but the degree of melting depends on the mechanism of metal descent, which might range from complete core merger during a giant accretionary impact, to emulsification and slow sinking of the core of a small planetesimal, to a rain of tiny metal droplets from an undifferentiated impactor \citep{Golabek+08, Samuel+Tackley08}.

Probably the main mechanism for generating a magma ocean from the entire silicate fraction of a planet or protoplanet is through giant collisions \citep{Tonks+Melosh93, Canup08, Nakajima+Stevenson15}. 
While giant collisions provide enough heat for whole-mantle melting, 
hemispheric or local magma oceans are also possible. 
\citet{Tonks+Melosh93} showed that if the isostatic rebound time is less than the freezing time of the magma, then the return of the planet's center of mass to the center of the body would result in extrusion of melt over the planet's surface as discussed in more detail by \citet{Reese+Solomatov06}.

\section{Partitioning in Accreting Planetary Embryos}
\label{sec:12-3}

During the main accretion phase of planetary embryos, 
a large amount of energy is brought into their interiors through accretionary impacts of planetesimals, as described above. 
If all the accretion energy is converted to the thermal energy without loss, 
the planetary embryo has an averaged interior temperature of the order of $10^4$~K, 
which is high enough that the planetary embryo is totally molten. 
However, 
since the cooling timescale is much shorter than the typical growth timescale of planetary embryos, 
the magma ocean is solidified immediately after formation. 
For magma oceans to be maintained, 
the presence of atmospheres is needed to delay the loss of energy from the planetary embryos.

Unless planetesimals are free of volatiles and grow in a vacuum, 
planetary embryos readily gain volatiles and have atmospheres, 
provided they become massive enough to bind the atmospheres gravitationally. 
The gravitational potential energy of an embryo of mass $M_e$ is $G M_e/ R_e$ = $G (4 \pi \bar{\rho}_e/3)^{1/3} M_e^{2/3}$, where $G$ is the gravitational constant and $R_e$ and $\bar{\rho}_e$ are the radius and mean density of the embryo, respectively, 
while the specific thermal energy of gas is $\sim$ $k T / (\mu m_\mathrm{H})$, where $k$ is the Boltzmann constant, $T$ is the temperature, $\mu$ is the mean molecular weight, and $m_\mathrm{H}$ is the mass of a hydrogen atom.
The condition is, thus, given with respect to embryo mass as 
$M_e >$ $\sim 6 \times 10^{22}$~$\mu^{-3/2}$~kg for a gas temperature of 300~K and a planetary mean density of 3500~kg/m$^3$.
This indicates  that an H$_2$O-vapor atmosphere is gravitationally bound on a planetary embryo of $\sim 8 \times 10^{20}$~kg, which is less massive than the moon. 
Even for an H$_2$ atmosphere, the lunar mass is large enough to attract it gravitationally, from the energetic point of view. 
The reason that the moon has no atmosphere today would be that it was subject to atmospheric erosion processes such as the solar wind-induced one.

Planetesimals, if containing water, release some fraction of their internal water upon impact, 
which results in forming a vapor atmosphere on an accreting planetary embryo in the following way. 
Impact-induced shockwaves propagate in the interior of the planetesimals. 
The shock heating is known experimentally to cause the internal water to evaporate off the planetesimals \citep[e.g.,][]{Lange+Ahrens82}. 
Since the impact velocity is almost the same as the escape velocity of the planetary embryo, 
it becomes high with increasing planetary embryo mass. 
When the embryo mass exceeds approximately the lunar mass, 
impact degassing starts to occur \citep[][]{Abe+Matsui85}. 
This type of atmosphere is termed an impact-generated steam atmosphere.

Since water vapor absorbs infra-red radiation well, 
the steam atmosphere traps the accretion energy of planetesimals and exerts a strong thermal blanketing effect. 
According to numerical models that simulate the accumulation and energy budget of the atmosphere and mantle simultaneously \citep[][]{Abe+Matsui85,Abe+Matsui86}, 
the impact-generated steam atmosphere on a Mars-mass embryo is thick enough to provide insulation that enables heat sources to cause mantle \ikoma{melting} and magma ocean \ikoma{formation}, as shown in Fig.~1. 
Provided the planetesimal accretion rate is high enough (see the solid and long-dashed lines in Fig.~1a), 
the surface temperature and atmospheric vapor mass are found not to increase monotonically, but to level off until just before the end of growth. 
As realized by \cite{Abe+Matsui86}, this coupled system of the accumulating atmosphere and molten mantle is a self-stabilizing system. 
This is because water is well mixed with molten silicate (i.e., magma). 
An increase of atmospheric vapor enhances the greenhouse effect, which leads to extending the molten area of the surface. 
This enhances dissolution of water into the magma ocean and thus decreases atmospheric vapor, 
leading to a reduction of the surface temperature. 
Consequently, because of this negative feedback loop, 
the surface temperature remains almost constant around the melting temperature of silicate ($\sim$~1500~K). 
This means that accreting embryos are always covered with magma oceans during the main accretion phase, 
provided the accretion rate is high enough.

\begin{figure}
	\begin{center}
	\includegraphics[width=1.0\textwidth]{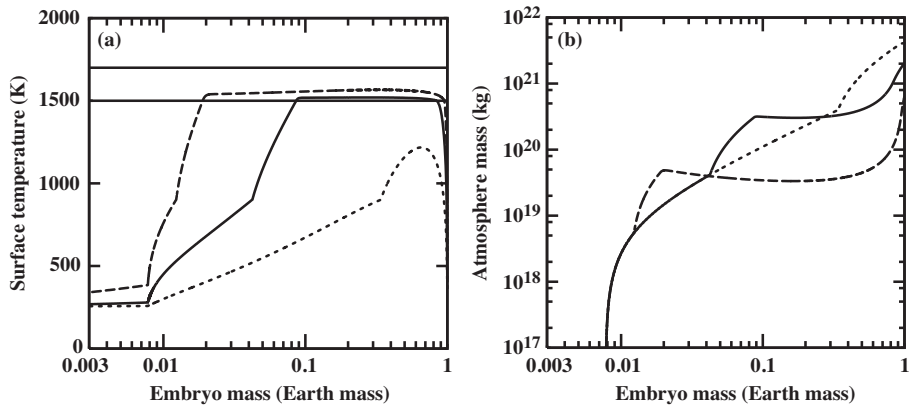}
	\caption{Growth of an impact-generated steam atmosphere. (a) The surface temperature and (b) atmospheric vapor mass of a growing planetary embryo are plotted as a function of planetary embryo mass. 
	The calculations have been done in a similar way to those of \cite{Abe+Matsui86}. 
	(Note that we have ignored the heat flux from the interior.)
	The solid line represents the growth with the fiducial value of planetesimal accretion rate adopted by \cite{Abe+Matsui86}, at which the embryo grows to the Earth mass in 50 million years. The long- and short-dashed lines represent the cases with 10 and 0.1 times the fiducial value, respectively. 
	The lower and upper horizontal solid lines in panel (a) indicate the assumed solidus and liquidus of silicate.}
	\end{center}
\label{fig-MI-1}
\end{figure}

Likewise, the mass of the steam atmosphere $M_a$ remains almost constant after the embryo mass exceeds 0.01-0.1 Earth masses, depending on the planetesimal accretion rate $\dot{M}_e$, as shown in Fig.~1b. 
This is interpreted as follows:
In a radiative atmosphere, the surface temperature $T_s$ is related to the energy flux $F$ and the optical thickness $\tau$ as $T_s^4 \propto F \tau$. 
In the present case, the energy flux results from planetesimal accretion and \ikoma{is}
written as $F \simeq G M_e \dot{M}_e / 4 \pi R_e^3$. 
Also, because the opacity $\kappa$ is assumed to be constant, 
the optical thickness is related to the pressure $P$ such that $\tau \simeq \kappa P / g $ $\simeq \kappa M_a / 4 \pi R_e^2$. 
Therefore, the surface temperature is related to the embryo mass, atmospheric mass, and planetesimal accretion timescale $t_\mathrm{acc}$ ($\equiv M_e/\dot{M}_e$) as 
\begin{equation}
	\sigma T_s^4 \simeq \frac{GM_e \dot{M}}{4 \pi R_e^3} \cdot \frac{\kappa M_a}{4 \pi R_e^2} 
	\propto \frac{M_e^{1/3} M_a}{t_\mathrm{acc}},
\end{equation}
or 
\begin{equation}
	M_a \propto \frac{t_\mathrm{acc}}{M_e^{1/3}} T_s^4.
\label{eq-MA86-vapormass}
\end{equation}
Since the surface temperature is fixed to the melting temperature of silicate, 
as found in Fig.~1a, the atmospheric mass is determined by the accretion timescale and the embryo mass. 
In the ordinary growth regime that \cite{Abe+Matsui86} assumed, $t_\mathrm{acc} \propto M_e^{1/3}$ \citep[e.g.,][]{Safronov72}, 
so that $M_a$ is independent of $M_e$ and constant through the growth. 
In the runaway growth regime, since $t_\mathrm{acc} \propto M_e^{-1/3}$ \citep[][]{Wetherill+Stewart89}, 
the atmospheric mass decreases with growth in such a way that $M_a \propto M_e^{-2/3}$. 
In any case, from the viewpoint of water partitioning, 
of particular importance would be the fact that all the accumulated water other than the atmospheric vapor is incorporated in the magma ocean and the core.

Hence, the water contents of planetesimals affect the amount of water that accreting planetary embryos finally contain in their interior, since the mass of the steam atmosphere is limited to more or less the mass of the current Earth's oceans ($\simeq$ 0.02~\% of the Earth mass), according to the self-stabilizing mechanism \ikoma{described} above. 
A systematic investigation of the impact-generated steam atmosphere was done by \citet{Zahnle+88}, 
who modeled the formation and growth of the atmosphere in a similar way to \citet{Abe+Matsui86}, 
but treated the atmospheric blanketing effect in more detail and included interchange of water and energy between the surface and interior, and hydrogen escape by the solar extreme ultraviolet ray (EUV) and by impact erosion. 
They confirmed that a wide range of the model parameters yield solutions of the \citet{Abe+Matsui86}-type atmosphere,  
which are favored by moderate ($\sim$ 0.1~\%) water content in the planetesimals. 
However, if the incoming planetesimals were too dry or the EUV flux too high, 
very little water would accumulate at the surface, which means the magma ocean is rather dry.

If embedded in a protoplanetary disk, planetary embryos attract gravitationally hydrogen and helium from the disk. 
This type of atmosphere is often termed the primordial atmosphere. 
The primordial atmosphere is also known to affect the water and energy budgets of planetary embryos. 
As shown by numerical models of the structure of the primordial atmosphere of an embedded planetary embryo \citep[][]{Hayashi+79, Ikoma+Genda06}, 
the atmospheric blanketing effect is strong enough for a magma ocean to be generated, 
if the embryo mass is larger than $\sim$ 0.3 Earth masses. 
This is a rather robust conclusion, because the surface temperature, which is given by 
\begin{equation}
	T_s \approx \frac{GM_e \mu \, m_\mathrm{H}}{4R_ek},
\label{eq-neb-temp}
\end{equation}
is independent of the opacity and energy flux (or accretion rate). 
In contrast to atmospheres of embryos in a vacuum, 
changes in those quantities are compensated by that in the atmospheric mass (i.e., the total optical thickness) in the case of the atmosphere connected to the surrounding gaseous disk. 
For example, if the planetesimal accretion rate decreases for some reason, 
the atmosphere contracts gravitationally, which leads to further accretion of gas. 
Analytically, the atmospheric mass is given by 
\begin{equation}
	M_a \approx \frac{\pi^2 \sigma}{3 \kappa L} \left(\frac{GM_e\mu \, m_\mathrm{H}}{k}\right)^4 
		\ln \left(\frac{R_\mathrm{B}}{R_e}\right)  
	\hspace{1ex}
	\propto 	\dot{M}_e^{-1} M_e^{10/3} \mu^4, 
\label{eq-neb-mass}
\end{equation}
where $L$ is the protoplanetary luminosity, which is given approximately by $GM_e \dot{M}_e / R_e$ in the accretion phase, $\sigma$ is the Stefan-Boltzman constant and $R_\mathrm{B}$ is the Bondi radius. 

As indicated by Eqs.~(\ref{eq-neb-temp}) and (\ref{eq-neb-mass}), 
the surface temperature and atmospheric mass depend strongly on the molecular weight $\mu$. 
The primordial atmosphere is usually assumed to have solar composition; 
namely, the atmosphere consists mainly of hydrogen and helium and contains only $\sim$ 0.1~\% H$_2$O. 
However, since oxides are available at the embryo surface, 
H$_2$O can be produced in the atmosphere through reactions with the atmospheric hydrogen. 
As discussed in \citet{Ikoma+Genda06}, 
for the H$_2$O production process to work efficiently, 
magma oceans should be maintained. 
In the case of the solar-composition atmosphere, 
the embryo mass must be larger than $\sim 0.3 M_\oplus$. 
The amount of H$_2$O thus produced depends on the oxidation state of the mantle. 
For example, the iron-w\"{u}stite (IW) buffer (1.894Fe + O$_2$ $\leftrightarrow$ 2 Fe$_{0.974}$O) has the ability of producing H$_2$O comparable in amount with H$_2$ \citep[][]{Rubie+78}.

\begin{figure}
	\begin{center}
	\includegraphics[width=1.0\textwidth]{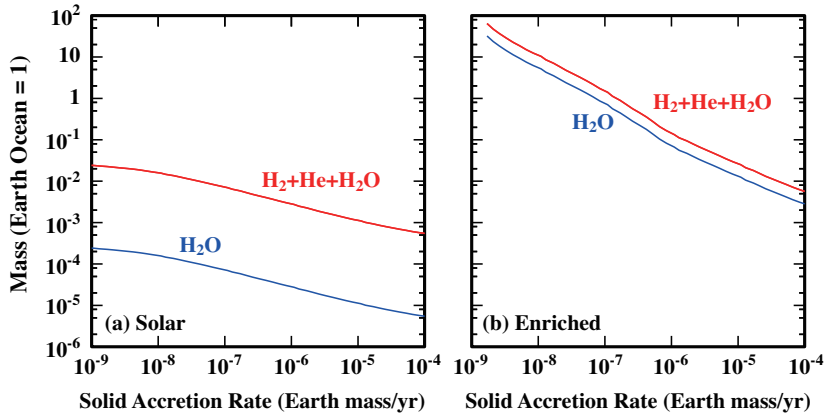}
	\caption{Mass of the primordial atmosphere on a Mars-mass embryo embedded in a protoplanetary disk, which is plotted as a function of planetesimal accretion rate. In panel~(a), the atmospheric composition is assumed to be solar. In panel~(b), the atmosphere is assumed to consist of 50\% H$_2$O and 50\% H$_2$+He. }
	\end{center}
\label{fig-MI-2}
\end{figure}

Figure~2 shows the mass of the solar abundance atmosphere (left panel) and the 50~\% water and 50~\% H/He mixed atmosphere (right panel) as functions of planetesimal accretion rate. 
Here the atmospheric structure has been integrated in a similar way to that of \cite{Venturini+15}. 
Comparing the two panels, one realizes that the water fraction has a great impact on the atmospheric mass. 
While the water fraction in the enriched atmosphere (panel b) is larger by a factor of $\sim$500 than that in the solar-abundance atmosphere (panel a), the difference in vapor mass is larger by about four orders of magnitude. 
In the case of $\dot{M}_e $ = $1 \times 10^{-7}$~$M_\oplus/{\rm yr}$, for example, 
the atmospheric H$_2$O is comparable in mass with the current Earth's oceans even for a Mars-mass embryo.

The thermal state of the mantle (e.g., the depth of the magma ocean) and the partitioning of the produced water between the atmosphere and magma ocean are yet to be investigated in the case of the embedded atmospheres, while those have been investigated in studies of impact-induced melting \cite[e.g.,][]{Vries+16} and core formaion \citep[e.g.,][]{Rubie+15b, Rubie+15}. 
As inferred from Eq.~(\ref{eq-neb-temp}), the surface temperature is more sensitive to the molecular weight of the atmospheric gas 
than the embryo mass. 
Thus, it is likely that the surface temperature of a Mars-mass embryo becomes higher than the melting temperature of rock, provided the atmosphere becomes enriched in water. 
The initial nebula-captured atmosphere of a Mars-mass embryo is not massive enough. However, 
if water is added to the atmosphere for some reason (for example, impact and evaporation of relatively wet materials), 
the atmosphere readily becomes massive, according to Eq.~(\ref{eq-neb-mass}).
Then, the surface temperature may exceed the melting temperature of rock, generating a magma ocean, as found from Eq.~(\ref{eq-neb-temp}). This means
the H$_2$O production mechanism starts functioning. 
Consequently, the captured atmosphere becomes quite massive even on a Mars-mass embryo.
Since this atmosphere contains a large amount of water, part of which is dissolved into the magma ocean, the mantle of planetary embryos may be much wetter than previously thought.

\section{Partitioning in \sakujo{Isolated} Protoplanets}
\label{sec:12-4}
The solidification and degassing processes of magma oceans set the initial interior and atmospheric volatile budgets for terrestrial planets. 
Since magma oceans do not degas perfectly, they provide a major mechanism for adding a small but potentially critical volatile fraction to planetary mantles; this volatile fraction will lower the melting temperature and viscosity of the mantle, encouraging the onset of thermal convection and volcanism. 
The interior and exterior volatile budgets are set by the degassing processes of the solidifying magma ocean.

Current models of magma ocean solidification mostly assume that volatiles degas whenever they are in excess of saturation, and demonstrate that even with nominal initial volatile concentrations, thick steam atmospheres are likely built above solidifying magma oceans, as discussed in the previous section \citep{Abe+Matsui85, Abe+Matsui86, Zahnle+88, Elkins-Tanton08, Lebrun+13}. 
Alternatively, degassing might be delayed if bubble nucleation requires a finite supersaturation pressure, as observed in the terrestrial context \citep{Bottinga+Javoy90, Lensky+06, Hardiagon+13}. 
Delayed and sudden degassing would imply supersaturated magma-ocean liquids that produce more volatile-rich mantles and comparatively late atmospheric formation. 
A later-forming atmosphere means faster early solidification through higher heat flux, unoccluded by a dense atmosphere. 
This process may also allow planets relatively closer to their stars to retain more volatiles in their interiors. Since stellar EUV radiation rapidly declines according to a power law in time, the later-forming atmosphere might avoid atmospheric escape.

\subsection{Magma ocean solidification}

A cooling protoplanet has four potential water reservoirs: an atmosphere, a magma ocean, a solid mantle, and a metallic core \citep{Elkins-Tanton12}. 
As magma ocean solidification proceeds, the water that terrestrial planets acquired during formation is partitioned among these reservoirs \citep[e.g.,][]{Zahnle+07,Elkins-Tanton08}. Out of these, the solid mantle is the least dynamic. Some of the water dissolved in the magma ocean can be incorporated into the solid mantle at the solidification front, but unless remelting occurs the water stored in the solid mantle has a minor influence on the evolution of the magma ocean and the atmosphere, which, on the other hand, are tightly coupled.

The liquid magma ocean is expected to contain some fraction of water (or hydroxyl) along with carbon and carbon compounds. Water, and almost all trace elements and volatiles, have extremely low solid/liquid partition coefficients and hence behave as incompatible elements on crystallization. Upon crystallization, water would thus be exsolved from solidifying cumulates and accumulate in the residual magma ocean liquids. As solidification proceeds, the magma ocean liquids become increasingly enriched in volatiles. When circulating at shallow depth, the volatiles degas into the growing atmosphere. 

At pressures and temperatures of magma ocean crystallization, no hydrous or carbonate minerals will crystallize \citep{Ohtani+04, Wyllie+Ryabchikov00}, although reducing conditions may stabilize graphite \citep{Hirschmann+Withers08}. Nonetheless, water and carbon dioxide will enter solidifying minerals in very small quantities. Even the small fractions of water or hydroxyl that are incorporated into the defects and voids of nominally anhydrous minerals can produce significant reductions in melting temperature and viscosity, and thus they can greatly influence the volcanic and dynamic evolution of planets. Olivine and orthopyroxene, for example, can contain in excess of 1000-1500 ppm of hydroxyl \citep[e.g.,][]{Hauri+06}. 
In general water or hydroxyl is partitioned into nominally anhydrous minerals at about 1:100 in relation to the water content of the magma, but exceptions such as ringwoodite exist \citep[see][supplementary material for compilations of mineral-melt partition coefficients]{Brown+14, Elkins-Tanton08}. 

Several groups have evaluated the storage rate of water, by considering the solid/liquid partitioning according to partition coefficients of water and its saturation limits for nominally anhydrous minerals, and/or by assuming the interstitial trapping of water-enriched melts \citep[][]{Elkins-Tanton08, Elkins-Tanton11,Lebrun+13,Hamano+13}. 
In some models, a constant mass fraction of 1~\% melt is assumed to be trapped in the solidifying cumulates over the solidification period. 
In these cases, the interstitial melts and the nominally anhydrous minerals contribute, respectively, about half the total interior reservoir \citep{Elkins-Tanton08}. 
This is because a bulk mineral-melt partition coefficient of water for lherzolite is about 1~\%, though the lack of experimental partitioning data makes quantitative estimation difficult under lower mantle conditions. 
The value of 1~\% for the mass fraction of trapped melts is comparable to the upper limit in the degree of melting estimated for abyssal peridotites at which melt separation starts \citep{Johnson+Dick92}. 
However, the efficiency of solid-liquid separation is highly uncertain at the bottom of the vigorously convective magma ocean, and possibly depends on the solidification rate itself.

The distribution of any elements that are incompatible in common mantle minerals is largely controlled by fractional crystallization, which in turn depends sensitively on whether growing mineral grains remain suspended or settle out \citep[][]{Solomatov+Stevenson93a, Solomatov+Stevenson93b, Solomatov+93, Solomatov00}. 
For fractional solidification to dominate, mineral grains must settle out and be effectively removed from equilibrium with the magma ocean liquid. 
As solids are removed, remaining liquids are progressively enriched in incompatible elements including radiogenic potassium and uranium, and the volatiles hydroxyl, nitrogen, and carbon. 
In batch solidification, liquids and crystals remain in equilibrium throughout solidification. 
These two crystallization sequences hence lead to completely different predictions for the dynamics of magma oceans, the relative importance of layering or mixing, as well as the mineral assemblage, bulk composition and trace-element composition of the solidified magma oceans. The relative importance of batch vs. fractional crystallization might explain the seemingly diverse signatures of the magma ocean on the Moon as compared to the relative scarcity of magma ocean signatures on the Earth \citep{Tonks+Melosh90, Solomatov+Stevenson93b}.

Quantifying when \ikoma{a} crystal settles or remains entrained in a solidifying magma ocean remains challenging. One line of reasoning argues that mineral grains are unable to settle if their settling velocity is small as compared to the convective flow speed in the magma ocean \citep{Tonks+Melosh90}. 
This rationale suggests that the characteristic crystal size is the most important factor for evaluating whether crystal settling occurs \citep{Solomatov+Stevenson93a, Solomatov+Stevenson93b, Solomatov+93, Solomatov00}. 
Analogue laboratory experiments, however, show that settling is possible along the boundaries of a rapidly convecting body even if the settling velocity of individual particles is small \citep{Martin+Nokes88}, but these experiments are limited to very low crystal fractions of $<$~1~\%. 

Recent direct numerical simulations of crystal settling in the lunar magma ocean \citep{Suckale+12b, Suckale+12a} suggest that the crystal fraction, defined as the local volume fraction of crystals, rather than the crystal size might be the determining factor for assessing crystal settling. The study shows that crystals may form a network at low crystal fractions of approximately 20~\%, which hinders the individual motion of crystals. Below this threshold, crystals decouple from the magmatic liquid, often in the form of clusters, and settle or rise, suggesting that fractional crystallization might be a valid approximation in this regime. Above that threshold, however, crystals are trapped in a network that moves with the ambient flow, which could be indicative of batch crystallization. An interesting ramification of this study \citep{Suckale+12b, Suckale+12a} is that the initial crystal fraction in the magma ocean might control its solidification history. If a magma ocean starts at a crystal fraction of 20\% because it melted incompletely, batch crystallization would dominate unless there is a significant remelting event. An entirely liquid magma ocean, however, could evolve through fractional crystallization initially and transition to a batch crystallization path later once solidification has progressed further. 
\begin{figure}
	\begin{center}
	\includegraphics[width=1.0\textwidth]{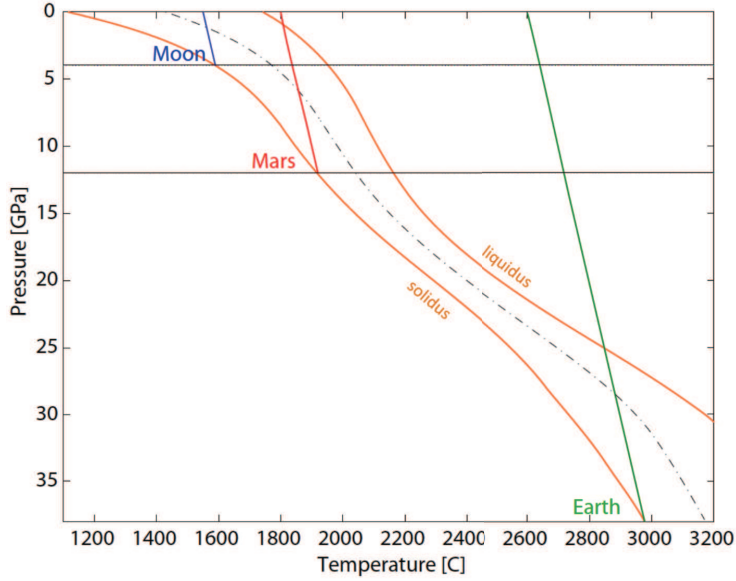}
	\caption{Idealized adiabats for the Moon (blue), Mars (red) and Earth (green) in comparison to the solidus and liquidus (both orange) for the mantle composition KLB-1, a mantle xenolith representative of fertile peridotite. 
	Correct adiabats deflect strongly toward the slope of the melting curves once they cross the solidus, but this figure still serves to convey the point that larger planets will have smaller depth ranges of solidification. 
	The plotted curves are fits to experimental data from \cite{Abe97}, \cite{Takahashi86}, \cite{Zerr+98}, and \cite{Tronnes+Frost02}. The grey horizontal lines indicate the pressure at the bottom of a magma ocean of approximately half the depth of the mantle maximum pressures in the mantle for Moon, Mars, and Earth based on planetary size. The potential temperatures for the Moon, Mars, and Earth were chosen such that their putative whole-mantle magma oceans would be just beginning to solidify, that is, the adiabat touches the solidus at their base.}
	\end{center}
\label{fig-JS-1}       
\end{figure}

Evaluating the crystal fraction of a magma ocean depends both on whether initial melting was complete and on the size of the planetary body. Figure~3 shows the solidi, liquidi, and adiabats for the Moon, Mars, and Earth. In all three cases, the adiabat first intersects the solidus at the bottom of the mantle, which suggests that solidification starts from below \citep{Walker+75}. Only on the Moon, however, does the adiabat fall between liquidus and solidus for almost the entirely solidification history. The steep slope of the adiabat for the depth range characteristic of the lunar mantle hence implies that crystals are stable at all depths. For large planets such as the Earth, only a relatively small portion of the magma ocean is likely to contain crystals for most of the solidification history as crystals may remelt when being convected upwards. Convection may thus not disperse crystals as easily as in a lunar magma \ikoma{ocean} implying that the crystal fraction in the partially molten depth range is likely high enough to limit or impede crystal settling. The role of crystals in determining the solidification history of the magma ocean could help understand the diverse signatures of magma oceans on the Moon as compared to Earth \citep{Tonks+Melosh90, Solomatov+Stevenson93b}. For example, if a terrestrial magma ocean extends to high pressures equivalent to the lowermost mantle on the Earth, energetics indicate that solidification may begin above the bottom of the magma ocean, creating a basal magma ocean solidifying downward \citep{Labrosse+07}.

Even complete fractional solidification will seldom drive water contents of the evolving magma ocean liquids high enough to create water-saturated minerals. Under the assumption in these models, to reach saturation in an Earth-sized planet, the magma ocean must begin with at least 10 wt\% of water \citep{Elkins-Tanton+Seager08}. 
Whether saturation is reached or not, later cumulates may form under higher oxygen fugacities than did early cumulates in a low-water magma ocean.

Late-solidifying minerals will have slightly higher trace and volatile element concentrations simply from partitioning out of an enriched fluid, but the incompatible element contents of the late-solidifying minerals will be tiny in comparison to the enriched interstitial fluids from the evolved magma ocean. \citet{Tonks+Melosh90} found that crystallization may be rapid in comparison to liquid percolation, so high fractions of retained liquid may be left in the cumulate mantle. The earliest solid mantle of the planet, then, would have a heterogeneous incompatible element composition. Interstitial melts and surface films will control the planet's internal budget of trace elements and volatiles, including hydroxyl, rare Earth elements, and noble gases.

Assuming that the solidifying mantle cumulates retain no interstitial liquids at all, \citet{Tikoo+Elkins-Tanton17} found that a terrestrial magma ocean that began with 0.5 wt\% of water produced a cumulate mantle with between a few and 45 ppm of water, enough to lower viscosity by a factor of 5 or so. The retention of interstitial liquids, however, can double the water content at the beginning of solidification (if, for example, 1\% of interstitial liquids containing 500 ppm water are retained), and can quadruple water content or more at the end of solidification (if 2\% of interstitial liquids containing 7,000 ppm are retained). Without going outside the range of water contents found in achondritic meteorites, these models can produce a hypothetical Earth with one or potentially multiple oceans' worth of water \citep{Elkins-Tanton11}.

\subsection{Magma-ocean degassing}

Previous models of magma-ocean solidification view degassing as an equilibrium process in which the volatile content in the magma ocean is assumed to follow solubility laws and the atmosphere grows continuously \citep{Elkins-Tanton08, Lebrun+13}. As volatile-rich magmatic liquid approaches the surface, water would be partitioned between the atmosphere and the magma ocean according to its solubility. Water could be released from the enriched magma to the atmosphere through two processes: nucleation and growth of bubbles and molecular diffusion of water across a thermal boundary layer. As long as the crystal fraction in the surface magma is sufficiently low for it to behave like a liquid (a soft magma ocean, defined by \citet{Abe97}), the bubble formation, growth, rise and burst of bubbles is likely the dominant process. 
It is difficult to quantify the exact value of the crystal fraction required, because it depends on a variety of parameters such as the crystal size distribution and crystal shape. For example, plagioclase-rich magmas can form load-bearing networks at crystal fractions of 10\%-20\% \citep{philpotts1999plagioclase} while other, more spherical minerals form networks at a typical rheologically critical crystal fraction of approximately 50\% \citep{renner2000rheologically}. 
Coalescence of bubbles during ascent could be fast enough to enable their efficient separation from the low viscous magma flow, by analogy to terrestrial volcanism \citep{Massol+16}.

If the high degree of supersaturation required for bubble nucleation, growth, and separation from magma flow is not present, then molecular diffusion is the dominant process and provides a minimum estimate for the degassing rate. 
The diffusion rate is controlled by gradient of the water content. 
As recently reviewed by \citet{Elkins-Tanton12}, the flow field in a molten magma ocean is sufficiently rapid to reach Rayleigh numbers in the range of 10$^{20}$-10$^{30}$. For comparison, a Rayleigh number on the order of 10$^7$ characterizes current solid-mantle convection on the Earth.
Based on scaling arguments \citep{Solomatov07}, the thermal boundary layer at the surface is  thin, on the order of cm, and the typical circulation time in the magma ocean ranges from a few days to less than one year. 
Thus, molecular diffusion may keep the atmosphere and the surface magma nearly at equilibrium \citep{Hamano+13}. 


Here and throughout the remainder of this chapter, we refer to continuous (equilibrium) degassing as a process where all volatiles in excess of their depth-dependent magma solubility degas instantaneously. Continuous degassing has been assumed in almost all existing magma ocean solidification models. Non-equilibrium processes are those that lead to a delay in degassing and imply a possible excess in volatiles at specific depths. 
Lacking observational constraints, it is currently not possible to conclusively determine the relative importance of the described continuous processes as opposed to non-equilibrium degassing in magma oceans. 
Here, we describe the latter mostly to complement the existing literature and to highlight potential observational signatures that could be indicative of non-equilibrium degassing. 
A variety of different factors could lead to non-equilibrium degassing. These include, but are not necessarily limited to, incomplete mixing of the magma ocean which could lead to an irregular distribution of volatiles horizontally and vertically, formation of a flotation crust or other solid-like surface features, incomplete melting of the magma ocean, likely most relevant if the magma ocean is created through impact cratering, a lack of mineral phases that allow for efficient nucleation of bubbles on crystal surfaces such as Fe-Ti oxides (heterogeneous nucleation), or high and/or variable crystal fraction in the magma ocean, particularly at shallow depth.

For whole-mantle magma oceans resulting from a significant melting event caused by a giant collision, we hypothesize that
the two primary factors that could lead to non-equilibrium degassing in magma oceans are 
(1) a lack of mineral phases that allow for efficient heterogeneous nucleation, which implies that a finite supersaturation pressure at depth is required for bubble nucleation and (2) incomplete escape of formed gas bubbles through a solidifying crystal mush. We briefly review existing constraints on both of these factors.

A melt that is supersaturated in volatiles exsolves a gas phase to reestablish thermodynamic and chemical equilibrium. 
Homogeneous nucleation theory \citep{Landau+Lifshitz80} attempts to quantify the energy required to nucleate bubbles spontaneously, but predicts unrealistically high supersaturation pressures in basaltic melts \citep{Sparks78, Sparks+94}. 
Bubble nucleation may be facilitated significantly by the existence of preferential nucleation sites such as surfactants, crystallites, or phenocrysts in the melt phase \citep{Blander+Katz75, Navon+Lyakhovsky98}. 
The supersaturation required for heterogeneous nucleation is determined by the surface energy of the crystal-gas as compared to the melt-gas interface \citep[e.g.,][]{Blander+Katz75} and is typically much lower than for homogeneous nucleation \citep{Hurwitz+Navon94, Gardner+Denis04}. 
An empirical criterion for comparing the two surface energies is the wetting angle θ. Unfortunately, the experimental determination of wetting angles is very challenging \citep{Gardner+Denis04}, but some constraints exist \citep{Eichelberger+82, Hurwitz+Navon94, Mangan+Sisson00}.

Experimental studies of the supersaturation required for bubble nucleation in basaltic liquids are rare \citep{Lensky+06,Hardiagon+13} and thus constraints for basaltic systems are often deduced from laboratory experiments performed in more silicic magma. 
Estimates for the supersaturation required for nucleation in rhyolitic magmas typically range from 120-200~MPa in melts exhibiting homogeneous or inefficient heterogeneous nucleation \citep{Mourtada-Bonnefoi99, Mangan+Sisson00, Mourtada-Bonnefoi02} to 5-20~MPa in systems where heterogeneous nucleation is very efficient \citep[e.g.,][]{Hurwitz+Navon94, Gardner+Denis04, Gardner07}. 
A similar span of supersaturation pressures may be expected for basaltic magma \citep{Bottinga+Javoy90} and is consistent with current estimates ranging from as little as 1-3~MPa \citep{Harris+Anderson83, Cashman+94}, to tens of MPa \citep{Edmonds+13}, 100-300~MPa \citep{Lensky+06} and even 1.5~GPa \citep{Hardiagon+13}. 
Summarizing, existing observational, experimental and theoretical evidence suggests the possibility of non-zero supersaturation pressures of tens of MPa and possibly higher in the absence of crystals with high wetting angles.

Once bubbles have been formed, they have to rise through the convecting magma ocean to burst and degas at the surface. The ability of bubbles to degas hence depends on the flow conditions in the vicinity of the surface, the presence of an atmosphere or conductive lid, and the crystallinity of the magma. It is interesting to note in this context that many natural examples of very efficient degassing on Earth refer to nearly aphyric mafic magmas mostly from Hawaii \citep{Harris+Anderson83, Cashman+94, Edmonds+13}. 

Degassing at a late stage of magma ocean solidification could be complicated due to a variety of factors. Due to the high solid-like viscosity, the circulation time of convection reaches up to several millions years \citep{Lebrun+13}. 
Also, a viscous thermal boundary layer starts to grow and its thickness reaches the order of 1-10 km, which could prevent the volatile exchange between the atmosphere and the surface magma. 
In this stage, melt percolation is a dominant process for melt extraction through the partially molten boundary layer \citep{McKenzie84, Solomatov07}. 
The melt percolation may carry up volatiles to the surface, but only if it occurs faster than the propagation of the solidification front \citep{Lebrun+13}. 

On the other hand, if fractional crystallization creates an unstable stratification due to compositional difference, a gravitational overturn can transport back and release volatiles into the deep interior \citep{Tikoo+Elkins-Tanton17}. If that is the case, volatiles might be transported back down into the mantle and not degas at all, leading to elevated volatile contents in portions of the solidified mantle after overturn. Another hypothetical consequence of degassing potentially occurring as late as just prior to or during overturn is that a finite bubble fraction could stabilize segments of the uppermost portion of the solidified magma ocean against overturn by reducing its density. This scenario, however, would require inefficient escape of the bubbles from the crystal mush as the time scales of solid-mantle overturn are much longer than those of gas percolation through a crystalline suspension. Since heterogeneous nucleation occurs preferentially on crystal interfaces, however, it is possible that a significant fraction of the bubbles could be trapped.
The chemical and thermal structure, and dynamics of a magma ocean in the late stages of solidification is particularly important to quantify the final amount of water retained in the interior.


\begin{figure}
	\begin{center}
	\includegraphics[width=1.05\textwidth]{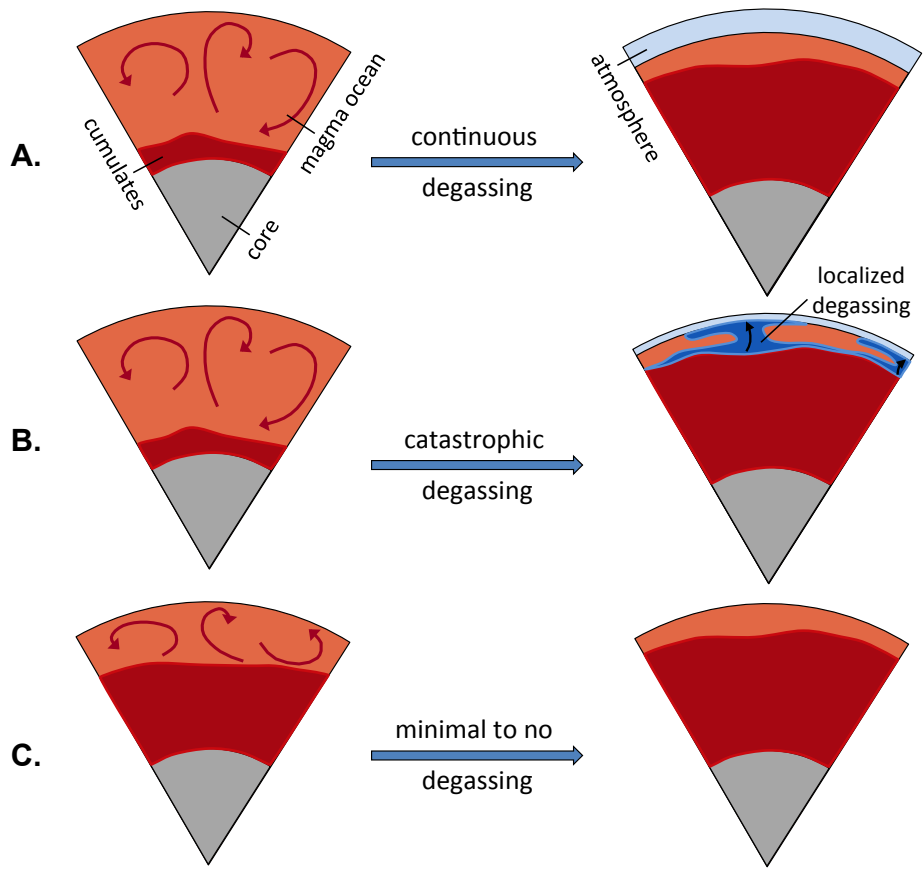}
	\caption{Schematic illustration of the three end-member cases of magma-ocean degassing.}
	\end{center}
\label{fig-JS-2}       
\end{figure}

Synthesizing, we suggest that degassing from solidifying magma oceans could be described by three end-member cases: (A) continuous degassing, (B) catastrophic degassing, and (C) negligible degassing (Figure~4).

\subsubsection*{(A) Continuous degassing}
Continuous degassing implies that degassing of volatiles in excess of saturation occurs constantly during the solidification of a magma ocean. This scenario is viable if heterogeneous nucleation requires negligible supersaturation. That might be the case, for example,  
if Fe-Ti oxides are abundant as small crystals in the liquid convecting magma, because oxides provide energy-efficient nucleation sites for bubbles \citep{Hurwitz+Navon94}. Alternatively, continuous degassing might arise if the volatile content is high enough to initiate degassing immediately after the formation of the magma ocean. 
It is also possible that the magma ocean may get continually replenished in volatiles by volatile-rich impactors, although impactors can play different roles depending on their size and the degree of melting upon impact and may lead to both volatile loss and replenishment. 

The existence of a steam-rich atmosphere that precedes magma ocean degassing would enhance degassing by keeping the surface of the magma ocean above its solidus, and eliminating the possibility of a solid crust \citep{Abe+Matsui86}. Without a solid layer, liquids can ascend to the thermal boundary layer at the planetary surface and experience the lowest possible pressures. They will thus degas more thoroughly because at the lowest pressures any given liquid has its supersaturation maximized.
The overall duration of the magma ocean phase would affect the total budget of water. 
As long as the magma ocean is sustained, all surface water is present in a vapor phase due to the hot surface temperature and therefore vulnerable to escape processes. 

We note that the mere existence of an atmosphere at the time of magma ocean formation, however, may not be sufficient to ensure continuous degassing. If partial pressures are low, the main effect of a primordial atmosphere could be to slow down the solidification process but degassing could still occur discontinuously. Magma oceans that degas continuously are characterized by substantial atmospheres and comparatively low solidification speed \citep{Zahnle+88, Elkins-Tanton08}.

\subsubsection*{(B) Catastrophic degassing} 
Delayed degassing could result if bubble nucleation required a finite supersaturation pressure of bubble trapping in the vicinity of the surface of the magma ocean. Even a relatively small supersaturation on the order of 10 MPa could result in delayed degassing, because it implies degassing occurring later in the solidification history when a larger portion of the magma ocean is already solidified or has a high crystal content. If finite supersaturation pressures are required for nucleation, it is possible that bubbles would form preferably in upwellings. In upwellings, compositional and thermal buoyancy reinforce each other and degassing should be possible essentially irrespective of bubble size. 

In the context of volcanic degassing on Earth, it has been noticed that high surface pressures could inhibit degassing rates notably and change the degassing composition \citep{gaillard2011atmospheric,gaillard2015redox}. Applied to degassing magma oceans, this finding would suggest that the presence of a substantial atmosphere with high surface pressures could reduce the efficiency of magma-ocean degassing, although thermal processes could have the opposite effect as discussed in the previous paragraph and it is difficult to know apriori how these competing processes would balance out.

If delayed degassing is related to bubble trapping, fluid and gas percolation is driven by the compaction of the cumulate pile. Degassing might hence occur preferentially in fluid-filled gas pathways. Both the preferential nucleation of bubbles in upwellings and the preferential nucleation of bubbles in fluid-filled pathways suggest that degassing occurs locally in flow regions that are conducive to nucleation events and may not be ubiquitous across the surface of the magma ocean. Delayed degassing could also entail a sequence of approximately simultaneous nucleation events throughout the magma ocean. 

Given that bubble nucleation and growth happen on time scales that are small compared to the time scale of magma ocean solidification, degassing may constitute a rapid and potentially even catastrophic event rather than a continuous process. The localized and sudden nature of late-stage degassing is an important difference compared to the more commonly invoked continuous degassing in excess of saturation \citep{Abe+Matsui85, Elkins-Tanton08, Lebrun+13}. 
Compared to continuous degassing, magma oceans that exhibit catastrophic degassing are characterized by a less substantial early atmosphere, higher volatile content in the solidified mantle, particularly the upper mantle, and a less homogeneous distribution of volatiles throughout the mantle. The composition of the atmosphere depends on the partitioning behavior of the volatiles involved \citep[e.g.,][]{Papale+06} and on the time scales of decompression relative to volatile diffusion. The faster the decompression, the less time is available for diffusion which could lead to substantial residual volatiles in the melt \citep{Mangan+Sisson00}.

\subsubsection*{(C) Negligible degassing} 
If the total volatile content is insufficient, a solidifying magma ocean may never reach the enrichment level required for degassing. Although volatile enrichment increases rapidly during the very late stages of solidification, very shallow magma oceans might not contain sufficient volatiles to reach the supersaturation pressure required for degassing or solid-mantle overturn may initiate earlier than degassing in which case a significant portion of the volatiles may be recycled into the mantle. Two magma oceans with comparable initial volatile contents but different depths might therefore experience drastically different degassing histories. Magma oceans with limited spatial extent on the planetary surface could also form a stable conductive lid, not unlike lava lakes, which hinders efficient degassing. The consequences of incomplete degassing for the planetary interior are substantial: Magma oceans that do not degas catastrophically will be characterized by elevated volatile contents during most of its solidification, which translates into higher volatile contents in the nominally anhydrous phases as well as in trapped interstitial liquid, minimal atmospheres and rapid time scales of solidification. A single deep magma ocean on a given planet might hence degas very differently than a sequence of shallow magma oceans \citep{Chambers13,Tucker+Mukhopadhyay14}.

The style of magma-ocean degassing could also depend on the types of volatiles. For example, water and carbon dioxide may play different roles in the degassing process, because of their differing solubility behavior at depth and diffusivities.

\subsection{Coupled Evolution of Magma Ocean and Atmosphere}
\label{sec:12-5}

The formation of an outgassed atmosphere affects the thermal evolution of a magma ocean. 
Without an atmosphere, the molten surface would emit significant thermal radiation, 
the flux of which is on the order of 10$^6$-10$^7$~W/m$^2$ due to high surface temperatures. 
As a result, the surface of an Earth-mass terrestrial planet would rapidly solidify in about 1000 years or less, even if a shallow, highly viscous magma ocean could continue to transport its heat effectively \citep{Solomatov07}. 
The greenhouse or blanketing effect of the outgassed atmosphere prevents heat from escaping from the surface. 
In general, as the atmosphere becomes massive, its greenhouse/blanketing effect reduces the outgoing radiation to space even further. 
Water vapor is one of the most potent greenhouse gases: 
Even 10 bar of a steam atmosphere, which is equivalent to about 1/3 of the total mass of the current Earth's oceans, reduces the outgoing radiation by about 2 orders of magnitude, compared with the corresponding blackbody radiation (see Fig.~5). 
$\mathrm{CO_2}$ is also important as an opacity source in the infrared, though its mass absorption coefficient, therefore its blanketing effect, is generally much smaller than that of water vapor \citep[e.g.][]{Marcq+17}.

The growth rate of the atmosphere would depend on how degassing proceeds (i.e,. equilibrium/non-equilibrium processes and volatile content in magmas), as described above. 
In the cases where super-saturation required for degassing becomes higher, degassing would occur in a more catastrophic way, leading to later formation of massive atmospheres. 
The outgassed atmosphere is also vulnerable to escape processes. 
Spectroscopic observations and atmospheric modelling of young stars have revealed that young, active stars emit higher X-ray and extreme ultraviolet radiation \citep[e.g.,][]{Ribas+05,Linsky+14,France+16}, which fuels hydrodynamic blow-off of early planetary atmospheres \citep[e.g.,][]{Watson+81,Kasting+Pollack83}. 
The hydrodynamic escape process could cause a significant loss of water from solidifying planets. 
This is because water vapor dissociates in the upper atmosphere by the intense EUV radiation, producing hydrogen atoms, which easily escape into space. 
This could affect the thermal evolution of a magma ocean, because the atmosphere grows or is depleted as a net result of the degassing and escape processes. 
Furthermore, 
the total amount of water lost during a magma ocean phase depends on its overall duration. As the magma ocean period becomes longer, the total inventory of water just after solidification becomes smaller.

Another possible process that affects the total water budget is impacts of planetary bodies. 
Terrestrial planets just after formation would experience a number of impacts by leftover fragments during accretion or planetesimals scattered from the outer orbital regions. 
These numerous impacts could erode or supply the atmosphere, depending on various parameters such as the impact velocity, the volatile content of impactors and their size distribution \citep[e.g.,][]{Chyba90, Vickery+Melosh90, deNiem+12}. As the net effect depends on the ambient atmospheric pressure as well, it could be worthwhile to examine the volatile budget by impacts along with the atmospheric growth during the magma ocean period in future work.

\begin{figure}
	\begin{center}
	\includegraphics[width=1.0\textwidth]{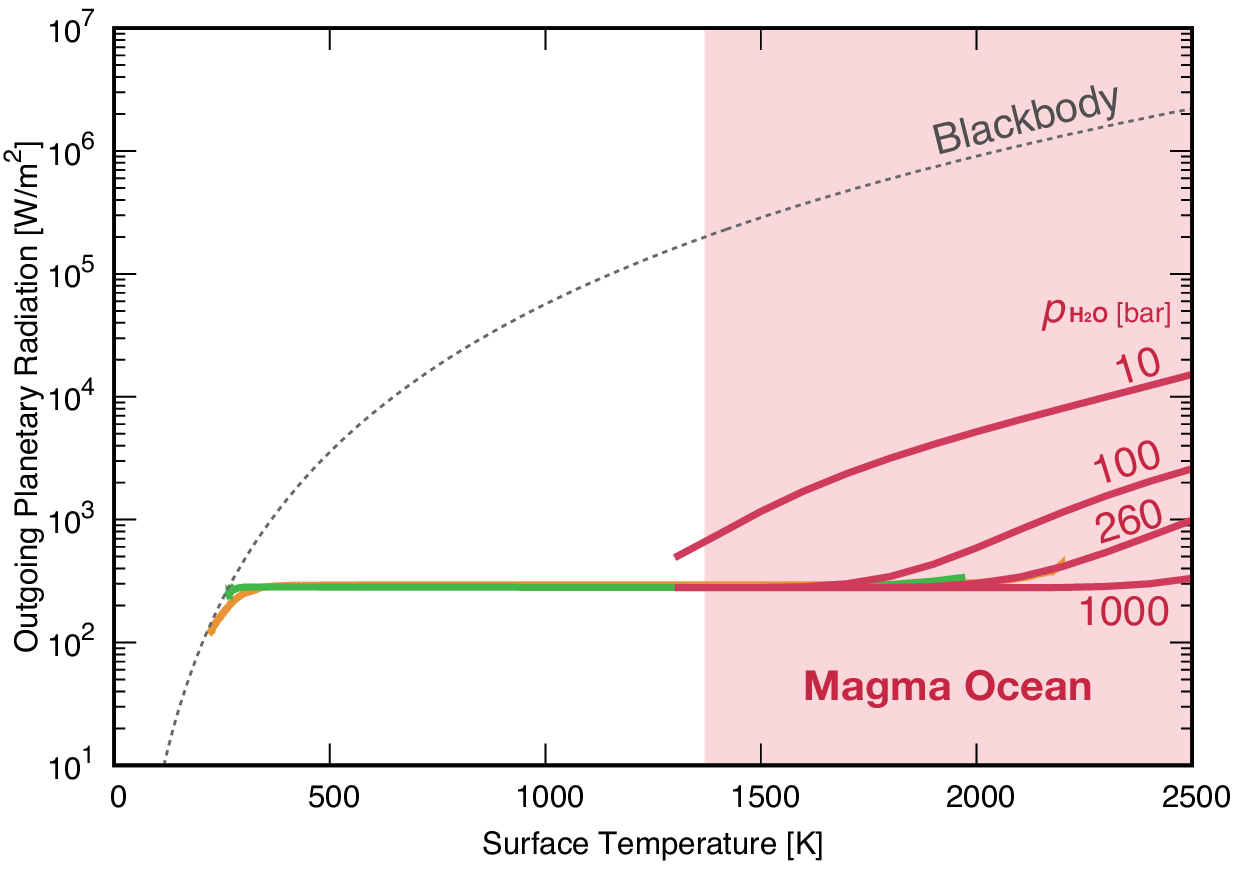}
	\caption{Thermal radiation as a function of surface temperature for various surface pressures of the steam atmosphere. Red lines are results obtained by using a radiative transfer model of \citet{Hamano+15}, compared with results by \citet{Goldblatt+13} (green) and \citet{Kopparapu+13} (orange) for a 260-bar steam atmosphere. 
	The outgoing radiation increases at the higher surface temperature for two reasons: (1) leakage of visible and near-infrared radiation from the surface and the hotter, lower part of the atmosphere, and (2) temperature increase at altitudes, where the optical depth measured from the top of the atmosphere is around unity.}
	\end{center}
\label{fig-KH-1}       
\end{figure}

Water partitioning in protoplanets has thus been examined with the thermal evolution of magma oceans, including feedback processes such as the greenhouse/blanketing effect of outgassed atmospheres, degassing of volatile species from the interior, \ikoma{and} loss of water associated with hydrodynamic escape of hydrogen \citep{Elkins-Tanton08, Lebrun+13, Hamano+13, Salvador+17}. 
In the case that steam-rich atmospheres are formed by degassing, terrestrial planets could have different thermal and water budget histor\ikoma{ies} during solidification, depending on their orbits \citep{Hamano+13, Lebrun+13}. 
As a result, the solidified planets would have two distinct characteristics: 
Planets that maintained certain amounts of water after their short solidification period, 
and others that dried out during a slow solidification process.

The evolutionary dichotomy originates from the presence of a lower limit in outgoing planetary radiation. 
As shown in Fig.~5, the thermal radiation from the top of the atmosphere decreases along with decrease in the surface temperature or increase in the atmospheric mass, but finally approaches $\sim$280~W/m$^2$\citep{Abe+88, Kasting88,Goldblatt+13, Kopparapu+13}. 
This means that the outgoing radiation from the top of steam atmospheres has a lower limit under high surface temperature conditions, such as a magma ocean phase. 
Hereafter, we referred to it as a radiation limit, which is also called a threshold for runaway greenhouse effects. 
The value of the radiation limit is determined by a saturation curve of water vapor, its opacity, and the planetary gravity \citep{Nakajima+92}. 
The value is less sensitive to including other gaseous species such as $\mathrm{CO_2}$ and $\mathrm{N_2}$, as long as water vapor is the most major component in the atmosphere \citep{Kopparapu+14, Marcq+17}.

Terrestrial planets receive the net stellar radiation, which could be smaller or larger than the radiation limit, depending on their orbits. 
According to \citet{Hamano+13, Hamano+15}, a critical orbital distance $a_\mathrm{cr}$ is defined as the distance from the host star where the net incoming stellar radiation is equal to the radiation limit,
\begin{equation}
	a_\mathrm{cr} \approx 0.83
	\left(\frac{F_\mathrm{lim}}{280~\mathrm{W \, m^{-2}}}\right)^{-1/2}
	\left(\frac{S_\ast (\tau_\mathrm{GI})}{0.7 S_\odot}\right)^{1/2}
	\left(\frac{1-\alpha_\mathrm{pl}}{1-0.2}\right)^{1/2}
	\, \mathrm{AU},
\end{equation}
where $F_\mathrm{lim}$ is the radiation limit, $\alpha_\mathrm{pl}$ is the planetary albedo, and $S_\odot$ and $S_\ast (\tau_\mathrm{GI})$ are, respectively, the bolometric fluxes at 1 AU of the current Sun and the host star at the time of giant impact. The critical distance separates the orbital regions of the two types of planet.

The presence of the radiation limit gives rise to a different water partitioning between the atmosphere and the interior, depending on the planetary orbit.
For a planet formed beyond the critical distance, 
as in the case of the Earth in Fig.~6(a), 
the primary reservoir of water after solidification is the steam atmosphere. 
More than half of the total inventory of water has already outgassed 
during the soft magma ocean stage.
After the crystal fraction at the surface reaches about 60~\%, the magma rheology abruptly changes due to a sudden jump of magma viscosity, and solid-state convection takes over from liquid-state convection \citep[a hard magma ocean, defined by][]{Abe97}. 
The final amount of water left in the interior would be established through this late stage of solidification, as discussed in the previous section. 
On the assumption of continuous (equilibrium) degassing in this later stage, the water stored in the early mantle accounts for less than 10\% of the total inventory, because most water has outgassed into the atmosphere.

Loss of water has little effect on the total water budget in this type of planet especially with a water content comparable to the total inventory of the present Earth. 
By definition, planets formed beyond the critical distance emit the outgoing radiation exceeding the incoming radiation, irrespective of the amounts of steam atmospheres. 
Consequently, Earth-sized planets solidify typically in less than several million years \citep{Lebrun+13, Hamano+13}, which is short enough for most of the primordial water to survive against the hydrodynamic escape. 
As long as the major component of the atmosphere is water vapor, the radiation limit also characterizes the cooling timescale of the atmosphere after solidification. 
In the case of a 300-bar steam atmosphere, water vapor would start to rain out on a timescale as short as about 1000 years after solidification, meaning that the surface water, which already outgassed by the time of complete solidification, would remain almost intact and contribute to the earliest water ocean. The following decrease in the mixing ratio of water vapor in the atmosphere would prevent further rapid loss of water \citep[e.g.][]{Kasting+Pollack83}. 

\begin{figure}
	\begin{center}
	\includegraphics[width=1.0\textwidth]{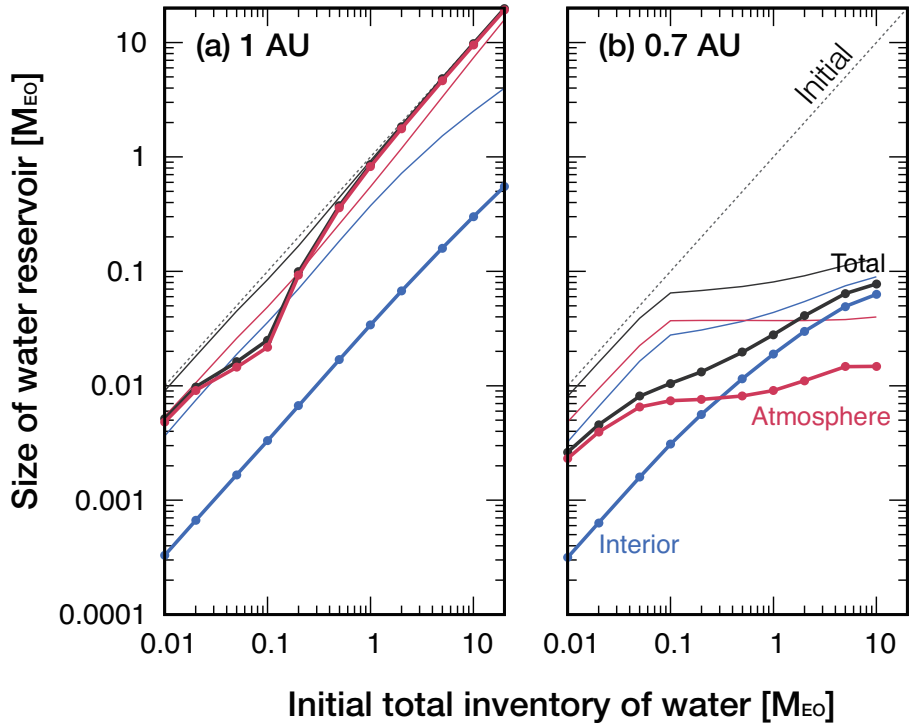}
\caption{
The bold lines indicate the relationship between the total inventory of water and the size of its reservoirs at the end of solidification of an Earth-sized planet at 1 AU(a) and 0.7 AU(b), on the assumption of solution equilibrium at the surface and with a mass fraction of trapped melt of 1~\%. 
The thin lines indicate those at the time that the potential temperature of the interior reaches 1700 K, at which the crystal fraction in the surface magma reaches 60~\%, a threshold for rheological transition. This corresponds to the transition from a soft magma ocean to a hard magma ocean. The reservoir size at potential temperature of 1700K would constitute a maximum estimate of the interior reservoir (the sum of solid mantle and magma ocean). All the results are obtained from \citet{Hamano+13}.}
	\end{center}
\label{fig-KH-2}       
\end{figure}

In contrast, if formed closer to the host star than the critical distance, a planet would become desiccated during crystallization by hydrodynamic escape. 
At orbits inside the critical distance, stellar radiation can keep the planetary surface in a molten state, if the planet has a steam atmosphere of about 30~bar or less, depending on the orbital distance (Fig.~5). 
In other words, the magma ocean would be sustained until the planet loses the steam atmosphere. 
The interior reservoir can account for a large fraction of the total inventory, especially in the case with the large initial total inventory of water. 
This is because the silicate melt has the higher water content in an early stage with the larger total inventory, whereas the amount of the steam atmosphere is regulated so that the net outgoing heat flux becomes positive. 

There have been arguments that dissociation of water molecules followed by preferential escape of hydrogen may cause a build-up of oxygen in the atmosphere, especially in the context of water loss from early Venus \citep[e.g.][and references therein]{Kasting95}. 
An extreme build-up of oxygen could possibly shut down the escape of hydrogen itself. This may not be the case during the solidifying phase of a magma ocean, because the dissociated oxygen atoms can be consumed by oxidation of ferrous iron in the abundant surface magmas \citep{Gillmann+09, Hamano+15}. As long as the magma ocean acts as an oxygen sink, this process would have a role in suppressing a significant build-up of abiotic oxygen on molten terrestrial planets \citep{Schaefer+16}. Recently, \citet{Marcq+17} have shown that addition of $\mathrm{CO_2}$ with a fixed amount of water vapor content reduces the blanketing effects. In the case \ikoma{of an} atmosphere consisting of 270 bar of $\mathrm{H_2O}$ and 90 bar of $\mathrm{CO_2}$, its blanketing effects would be still enough for keeping a long-lived magma ocean on early Venus, as long as the $\mathrm{CO_2}$-$\mathrm{H_2O}$ atmosphere have the albedo as low as that of a thick $\mathrm{H_2O}$ atmosphere ($\sim$ 0.2 for the Sun). However, if subsequent cloud formation enhances the planetary albedo, early Venus could have started to quickly solidify and water ocean could have formed, depending on the amount of water remaining in the atmosphere \citep{Salvador+17}. If a rotation rate of early Venus was as low as the present one, formation of high thick clouds on the day side might have helped to reflect \ikoma{a large} fraction of the incident sunlight \citep{Yang+14,Way+16}. On the other hand, clouds also have a role to enhance the blanketing effects. It would be necessary to investigate a balance of the opposing effects by clouds on planetary albedo and a blanketing effect to address early history of Venus.

The amount of water left in the interior increases with the assumed mass fraction of trapped melts. 
With a trapped melt fraction of 1~\%, the interior reservoir is the primary water reservoir in the case of the initial water inventory comparable to the total water inventory on the present Earth, 1.25-5 times the current ocean mass \citep[e.g.,][]{Hirschmann06}, as shown in Fig.~6. 
Note that solution equilibrium between the atmosphere and the magma ocean is assumed in these calculations. 
If degassing requires some degree of super-saturation of water in silicate melts,  it leads to the later formation of the outgassed atmosphere. As a result, the solidification time of the magma ocean would become shorter. 
In addition, the silicate melts keep the higher volatile content during solidification, which leads to the formation of a wetter early mantle. 
According to calculations using a 50-fold water solubility, which are intended to mimic the effects of super-saturation of water, 
the amount of water remained in the interior increases by a factor of 4 at most, compared to the results with a nominal water solubility \citep{Hamano+13}.

These conclusions may be qualitatively extended to terrestrial rocky planets, ranging from Mars-sized to super-Earths, as long as heat flux from a magma ocean is limited by radiative balance at the surface. 
The value of the radiation limit slightly increases with planetary mass, and is about 20~\% larger for 5-Earth-mass planets, compared to that for Earth-mass planets \citep{Kopparapu+14}. 
In this case, the critical distance from the star becomes about 10~\% smaller. 
Experimental and ab initio studies of melting curves are still absent at high pressures above 150~GPa.
In the case that super-Earths start to solidify with higher potential temperature\ikoma{s} compared to the Earth, 
magma oceans of super-Earths could be capped by silicate atmospheres at early stages of solidification. 
Since the planet would emit outgoing radiation as high as 10$^6$~W/m$^2$ from the top of hot silicate clouds, this phase lasts for at most several thousand years for 5-Earth-mass planets. 
The atmospheric escape thus would not affect the total budget of water, but some water might be incorporated into cumulates during this short phase, depending on the efficiency of melt/solid separation. 
On the assumption that water in excess of the saturation limit is degassed from the surface, larger planets can retain larger fractions of water in the interior for a given bulk content of water \citep{Elkins-Tanton11}. 
In the case of a 5-Earth-mass rocky planet formed beyond the critical distance, the fraction of water retained in the final 1~\% of silicate melt increases by a factor of less than 2, compared to that of an Earth-mass planet.

\section{Concluding remarks}
\label{sec:12-6}

Not only the presence of oceans, but also the current  
mass of ocean water is essential  
for making the Earth habitable, as described in \ikoma{the} Introduction. 
In this chapter, however, one has found it really challenging to answer the question of what determined the present total mass of the oceans on the Earth.  
Quite uncertain are how water/ice was distributed in the solar nebula and which processes dominated water delivery to the Earth. 
In addition, while magma oceans are readily formed, 
water partitioning  
between the atmosphere, magma ocean, solid mantle \ikoma{and core} 
in accreting planetary embryos and  
protoplanets is complicated and poorly understood.
The partitioning of water is also affected significantly by its accumulation and also mutually coupled with its escape. 
Processes relevant to water partitioning that are yet to be understood well include the solubility of water in peridotite liquid, 
the distribution of volatiles during crystallization, crystal settling and entrainment in solidifying magma oceans, non-equilibrium degassing such as bubble formation, and so on.

The initial amounts of water in planetary building blocks are obviously a key that determines the final amount of ocean water on terrestrial planets. 
As mentioned in Sect.~\ref{sec:12-2}, we are undergoing a paradigm shift regarding planet formation. 
Detection of many exoplanets with short orbital periods has compelled us to accept a dynamic picture including planetary migration and radial drift of solids, instead of the conventional, static picture. 
Also, \ikoma{rather than} km-sized planetesimals, cm-sized pebbles may contribute dominantly to planetary accretion. 
If that is the case, since pebbles come from beyond the snowline in general and thus contain large amounts of water, water delivery and accumulation processes differ greatly from those previously considered. 
Such a great degree of freedom in planetary formation theories demands much more observational constraints. 
Fortunately, new observational constraints are expected to come from both exoplanet observations and solar system explorations in the coming years.

The first-generation exoplanet survey projects, \textit{Kepler} and CoRoT, already confirmed that planetary systems are common. 
At present, the second-generation survey projects such as K2, TESS, and PLATO are under way or scheduled for detecting exoplanets orbiting nearby stars, which are suitable for further characterization. 
Those planets will include terrestrial planets with different amounts of water. 
The atmospheres of those planets will be characterized mainly via transit spectroscopy in the optical to infra-red by the upcoming space telescope JWST and the planned space telescope ARIEL. 
Key species such as water, methane, and oxygen will be hopefully detected in those atmospheres. 
Also, although being too close to host stars to be habitable, super-Earths covered with magma oceans might be detected \citep{Ito+15}. 
Further characterization of Earth-like planets will be challenged by the 30m-class ground-based telescopes E-ELT and TMT. 
With those telescopes, Earth-like planets covered with magma oceans might be detected in the future \citep{Lupu+14, Hamano+15}. 
For the details, see Chapter~7 in this volume (Noack et al. 2017).

The initial atmospheres of planets are determined largely by the material that accretes to create the planet. 
Small bodies in our solar system (in addition to meteorites) are the observable targets from which we learn more. In the wake of the highly successful Rosetta and Dawn missions, which visited a comet and the asteroids Vesta and Ceres, future missions will examine additional suites of differentiated and primitive rocky bodies. The OSIRIS-REx and Hayabusa2 missions are on their ways to the carbonaceous chondrite near-Earth asteroids Bennu and Ryugu, respectively, and will return surface samples to the Earth. In 2021 the Lucy mission will launch to the Jupiter Trojans, and will send back compositional data about these little-known but presumably ice-rich remnants of planet-building. In 2022 the Psyche mission to the metal asteroid Psyche will be launched, and will examine there what is hypothesized to be the core of a planetesimal, stripped of its silicate mantle in the first millions of years of the solar system. Through data from these missions we will be able to further constrain the starting compositions of growing planets.

%
%

\begin{acknowledgements}
First of all, we would like to express our deepest appreciation to Prof. Yutaka Abe, who passed away recently. We learned a lot from discussion with him, without which we could not complete this chapter. 
We thank anonymous referees for their careful reading and constructive comments, which helped us improve this manuscript greatly. 
Also we thank the SOC members of the ISSI workshop ``The Delivery of Water to Protoplanets, Planets and Satellites" for giving us the opportunity to write this chapter. 
M.~I. is supported by the Astrobiology Center Program of National Institutes of Natural Sciences (NINS) (No.~AB291004) and JSPS Core-to-Core Program ``International Network of Planetary Sciences”. 
K.H. acknowledges JSPS KAKENHI Grant Numbers JP26800242 and JP16J06133, and MEXT KAKENHI Grant Number JP17H06457.
\end{acknowledgements}

\bibliographystyle{aps-nameyear}      
\bibliography{refs}                
\nocite{*}

\end{document}